\documentclass[aps,preprint]{revtex4}%
\usepackage{amsfonts}
\usepackage{amsmath}
\usepackage{amssymb}
\usepackage{graphicx}%
\begin{document}
\preprint{CTP-SCU/2015009}
\title{Black Hole Radiation with Modified Dispersion Relation in Tunneling Paradigm:
Free-fall Frame}
\author{Peng Wang}
\email{pengw@scu.edu.cn}
\author{Haitang Yang}
\email{hyanga@scu.edu.cn}
\author{Shuxuan Ying}
\email{2013222020007@stu.scu.edu.cn}
\affiliation{Center for Theoretical Physics, College of Physical Science and Technology,
Sichuan University, Chengdu, 610064, PR China}

\begin{abstract}
Due to the exponential high gravitational red shift near the event horizon of
a black hole, it might appear that the Hawking radiation would be highly
sensitive to some unknown high energy\ physics. To study effects of any
unknown physics at the Planck scale on the Hawking radiation, the dispersive
field theory models have been proposed, which are variations of Unruh's sonic
black hole analogy. In this paper, we use the Hamilton-Jacobi method to
investigate the dispersive field theory models. The preferred frame is the
free-fall frame of the black hole. The dispersion relation adopted agrees with
the relativistic one at low energy but is modified near the Planck mass
$m_{p}$. The corrections to the Hawking temperature are calculated for massive
and charged particles to $\mathcal{O}\left(  m_{p}^{-2}\right)  $ and neutral
and massless particles with $\lambda=0$ to all orders. The Hawking temperature
of radiation agrees with the standard one at the leading order. After the
spectrum of radiation near the horizon is obtained, we use the brick wall
model to compute the thermal entropy of a massless scalar field near the
horizon of a 4D spherically symmetric black hole and a 2D one. Finally, the
luminosity of a Schwarzschild black hole is calculated by using the geometric
optics approximation.

\end{abstract}
\maketitle
\tableofcontents

%\affiliation{Center for Theoretical Physics, College of Physical Science and Technology,
%Sichuan University, Chengdu, 610064, PR China}

%\affiliation{Center for Theoretical Physics, College of Physical Science and Technology,
%Sichuan University, Chengdu, 610064, PR China}

\section{Introduction}

Soon after Stephen Hawking demonstrated that quantum effects could allow black
holes to radiate a thermal flux of quantum particles\cite{IN-Hawking:1974sw},
it was realized that there was the trans-Planckian problem with the
calculation\cite{IN-Unruh:1976db}. Hawking radiation appears to come from the
modes with huge initial frequencies, well beyond the Planck mass $m_{p}$,
which experience exponential high gravitational red-shifting near the horizon.
So the Hawking radiation relies on the validity of quantum field theory in
curved spacetime to arbitrary high energies. On the other hand, quantum field
theory is considered more like an effective field theory of an underlying
theory whose nature remains unknown. This observation poses the question of
whether any unknown physics at the Planck scale could strongly influence the
Hawking radiation.

To study the trans-Planckian problem, a hydrodynamic analogue of a black hole
radiation was considered\cite{IN-Unruh:1994je}. Following the Unruh's work,
there have been a lot of studies to understand the dispersive field theory
models\cite{IN-Unruh:1994je,IN-Brout:1995wp,IN-Corley:1996ar,IN-Corley:1997pr,IN-Himemoto:1999kd,IN-Saida:1999ap,IN-Unruh:2004zk,IN-Macher:2009tw,IN-Coutant:2011in,IN-Coutant:2014cwa,IN-Belgiorno:2014bna}%
, which focused on studying the effect on the Hawking radiation due to
modifications of the dispersion relations of matter fields at high energies.
Similar to the original method for deriving the Hawking radiation, the energy
fluxes for outgoing radiation were usually obtained by calculating the
Bogoliubov transformations between the initial and final states of incoming
and outgoing radiation. In most works, the Hawking effect could be recovered
at leading order\ under some suitable assumptions, which have been briefly
reviewed in \cite{IN-Barcelo:2005fc,IN-Jacobson:1999zk}.

After the Hawking's original derivation, there have been some other methods
proposed to understand the Hawking radiation. Recently, a semiclassical method
of modeling Hawking radiation as a tunneling process has been developed and
attracted a lot of attention. This method was first proposed by Kraus and
Wilczek\cite{IN-Kraus:1994by,IN-Kraus:1994fj}, which is known as the null
geodesic method. They employed the dynamical geometry approach to calculate
the imaginary part of the action for the tunneling process of s-wave emission
across the horizon and related it to the Hawking temperature. Later, the
tunneling behaviors of particles were investigated using the Hamilton-Jacobi
method\cite{IN-Srinivasan:1998ty,IN-Angheben:2005rm,IN-Kerner:2006vu}. In the
Hamilton-Jacobi method, one ignores the self-gravitation of emitted particles
and assumes that its action satisfies the relativistic Hamilton-Jacobi
equation. The tunneling probability for the classically forbidden trajectory
from inside to outside the horizon is obtained by using the Hamilton-Jacobi
equation to calculate the imaginary part of the action for the tunneling
process. Using the null geodesic method and Hamilton-Jacobi method, much fruit
has been
achieved\cite{IN-Hemming:2001we,IN-Medved:2002zj,IN-Vagenas:2001rm,IN-Arzano:2005rs,IN-Wu:2006pz,IN-Nadalini:2005xp,IN-Chatterjee:2007hc,IN-Akhmedova:2008dz,IN-Akhmedov:2008ru,IN-Akhmedova:2008au,IN-Banerjee:2008ry,IN-Singleton:2010gz}%
. Furthermore, the effects of quantum gravity on the Hawking radiation have
been discussed in the Hamilton-Jacobi method. In fact, the minimal length
deformed Hamilton-Jacobi equation for fermions in curved spacetime have been
introduced and the modified Hawking temperatures have been
derived\cite{IN-Chen:2013pra,IN-Chen:2013tha,IN-Chen:2013ssa,IN-Chen:2014xsa,IN-Chen:2014xgj,IN-Mu:2015qta}%
. These have motivated us to use the Hamilton-Jacobi method to study the
dispersive field theory models\cite{SF}. In this paper, we focus on the
dispersive models with the free-fall preferred fame, whereas those with the
static preferred fame have been studied in \cite{SF}.

The remainder of our paper is organized as follows. In section \ref{Sec:DHJM},
the deformed Hamilton-Jacobi equations are derived for the dispersive models
with the free-fall preferred frame. We then solve the deformed Hamilton-Jacobi
equations to obtain tunneling rates for massive and charged particles to
$\mathcal{O}\left(  m_{p}^{-2}\right)  $ and massless and neutral particles to
all orders. The thermal entropy of a massless scalar field near the horizon is
computed in section \ref{Sec:EBWM} using the brick wall model. In section
\ref{Sec:BHE}, we calculate the luminosity of a Schwarzschild black hole with
the mass $M\gg m_{p}$. Section \ref{Sec:Con} is devoted to our conclusion.
Throughout the paper we take Geometrized units $c=G=1$, where the Planck
constant $\hbar$ is square of the Planck mass $m_{p}$.

\section{Deformed Hamilton-Jacobi Method}

\label{Sec:DHJM}

In this section, we first derive the deformed Hamilton-Jacobi equation
incorporating the modified dispersion relation (MDR) assuming that the
preferred reference frame is the free-fall frame. We then solve the deformed
Hamilton-Jacobi equation for the imaginary part of $I$ which gives the
tunneling rate $\Gamma$ across the event horizon. We consider two cases, a
massive and charged particle to $\mathcal{O}\left(  m_{p}^{-2}\right)  $ and a
neutral and massless particle with $\lambda=0$ to all orders.

\subsection{Deformed Hamilton-Jacobi Equation}

To study the deformed Hamilton-Jacobi method incorporating the MDR for the
Hawking radiation, one first needs to choose the form of MDR in flat spacetime
(the local free-fall frame) and generalizes it to curved spacetime. To be as
general as possible, we will work with the MDR for a particle with mass $m$
\begin{equation}
E^{2}=F^{2}\left(  p\right)  +m^{2},\label{eq:MDR}%
\end{equation}
where we define%
\begin{equation}
F\left(  p\right)  =p%
%TCIMACRO{\dsum \limits_{n=0}}%
%BeginExpansion
{\displaystyle\sum\limits_{n=0}}
%EndExpansion
C_{n}\frac{p^{2n}}{m_{p}^{2n}},\label{eq:F(p)}%
\end{equation}
$m_{p}$ is the Planck mass, $C_{0}=1$, and $E$ and $p$ are the energy and the
norm of the momentum measured in some preferred reference frame, respectively.
Note that the MDR $\left(  \ref{eq:MDR}\right)  $ is rotational invariant in
4D spacetime. To generalize the MDR $\left(  \ref{eq:MDR}\right)  $ to curved
spacetime with the metric $g_{\mu\nu}$, we denote by $u^{\mu}$ the unit vector
field tangent to the observers' world lines, which picks up a preferred frame.
For a particle with the energy-momentum vector $p_{\mu}$, the energy $E$ and
the norm of the momentum $p$ of the particle measured by these observers are%
\begin{align}
E &  =p_{\mu}u^{\mu},\label{eq:energy}\\
p^{2} &  =E^{2}-p_{\mu}p^{\mu}.\label{eq:momentum}%
\end{align}
The curved spacetime generalization of the MDR $\left(  \ref{eq:MDR}\right)  $
with a preferred frame described by $u^{\mu}$ is obtained by plugging eqns.
$\left(  \ref{eq:energy}\right)  $ and $\left(  \ref{eq:momentum}\right)  $
into eqn. $\left(  \ref{eq:MDR}\right)  $. To obtain the deformed
Hamilton-Jacobi equation incorporating the MDR, one needs to relate the
classical action $I$ to $p_{\mu}$. In fact, it can be shown that, if $I$ is a
solution of the Hamilton-Jacobi equation, then the transformation equations
give%
\begin{equation}
p_{\mu}=-\partial_{\mu}I,\label{eq:transformation-eqn}%
\end{equation}
where $-$ appears since $p_{\mu}=\left(  E,-\vec{p}\right)  $ in our metric
signature. Replacing $p_{\mu}$ with $I$ via eqn. $\left(
\ref{eq:transformation-eqn}\right)  $ and putting eqns. $\left(
\ref{eq:energy}\right)  $ and $\left(  \ref{eq:momentum}\right)  $ into eqn.
$\left(  \ref{eq:MDR}\right)  $ gives the deformed Hamilton-Jacobi equation.
In the appendix of \cite{SF}, the deformed Hamilton-Jacobi equation is also
derived in the language of the effective field theory for a scalar field and a
fermion one. There we considered a scalar/fermion with the mass $m$ and charge
$q$ in a static black hole with the presence of electromagnetic potential
$A_{\mu}$. Neglecting self-interacting effective operators, we constructed the
$U\left(  1\right)  $ gauge invariant effective field theory incorporating the
MDR. The deformed Klein--Gordon/Dirac equation was derived. The deformed
Hamilton-Jacobi equation for scalars/fermions was then obtained using the WKB
approximation. It was found there that the deformed scalar/fermionic
Hamilton-Jacobi equation with respect to the preferred frame $u^{\mu}$ in the
black hole background spacetime can be written as%
\begin{equation}
T^{2}=F^{2}\left(  X\right)  +m^{2},\label{eq:Deformed-HJ}%
\end{equation}
where%
\begin{equation}
T=-u^{\mu}\left(  \partial_{\mu}I+qA_{\mu}\right)  ,\text{ }X^{2}%
=T^{2}-\left(  \partial_{\mu}I+qA_{\mu}\right)  ^{2},\label{eq:T and X}%
\end{equation}
$A_{\mu\text{ }}$is the black hole's electromagnetic potential and $q$ is the
particle's charge.

As in \cite{SF}, we here consider the black hole whose metric in the
Schwarzschild-like coordinate is given by
\begin{equation}
ds^{2}=f\left(  r\right)  dt^{2}-\frac{dr^{2}}{f\left(  r\right)  }-C\left(
r^{2}\right)  h_{ab}\left(  x\right)  dx^{a}dx^{b},\label{eq:SW-Coordinate}%
\end{equation}
where $f\left(  r\right)  $ has a simple zero at $r=r_{h}$ with $f^{\prime
}\left(  r_{h}\right)  $ being finite and nonzero. The vanishing of $f\left(
r\right)  $ at point $r=r_{h}$ indicates the presence of an event horizon. We
also assumed that the black hole is asymptotically flat which gives $f\left(
r\right)  \rightarrow1$ as $r\rightarrow\infty$. However, a more suitable
coordinate for describing a specific family of freely falling observers is the
Painleve-Gullstrand (PG)
coordinate\cite{IN-Corley:1996ar,IN-Corley:1997pr,DHJM-Corley:1997ef}. The PG
coordinate anchored to the freely falling observers along the radial direction
takes the form of%
\begin{equation}
ds^{2}=dt_{p}^{2}-\left[  dr-v\left(  r\right)  dt_{p}\right]  ^{2}-C\left(
r^{2}\right)  h_{ab}\left(  x\right)  dx^{a}dx^{b},\label{eq:PE-Coprdinate}%
\end{equation}
where $v\left(  r\right)  $ is the velocity of the free fall observer with
respect to the rest observer and $t_{p}$ measures proper time along them. The
spacetime also has the event horizon at $r_{h}$ satisfying $v\left(
r_{h}\right)  =-1$. We assume $v<0,$ $dv/dr>0$ and $v\rightarrow v_{0}\leq0$
as $r\rightarrow\infty$. Note that $v<0$ means the infalling observers. For
simplicity we specialize to the particular family of observers with $v_{0}=0$
who start at infinity with a zero initial velocity. Since the vector field
$u^{\mu}$ of the freely falling observers is tangent to the infalling world
lines, one has for the infalling observers along the radial direction with
$v_{0}=0$ that
\begin{equation}
u^{\mu}=\left(  1,v\left(  r\right)  ,\vec{0}\right)  \label{eq:U-PE}%
\end{equation}
in the PG coordinate and%
\begin{equation}
u^{\mu}=\left(  \frac{1}{f\left(  r\right)  },\sqrt{1-f\left(  r\right)
},\vec{0}\right)  \label{eq:U-SW}%
\end{equation}
in the Schwarzschild-like coordinate. The fact that $t_{p}$ is the proper time
along the infalling world lines means that $u_{\mu}$ is equal to the gradient
of $t_{p}$,
\begin{equation}
u_{\mu}=\partial_{\mu}t_{p}.\label{eq:U-Mu}%
\end{equation}
Using eqns. $\left(  \ref{eq:U-SW}\right)  $ and $\left(  \ref{eq:U-Mu}%
\right)  $ gives
\begin{equation}
t_{p}=t+\int\frac{\sqrt{1-f\left(  r\right)  }}{f\left(  r\right)
}dr.\label{eq:Tp-T}%
\end{equation}
Substituting eqn. $\left(  \ref{eq:Tp-T}\right)  $ into the PG metric $\left(
\ref{eq:PE-Coprdinate}\right)  $ and comparing it to the Schwarzschild-like
coordinate $\left(  \ref{eq:SW-Coordinate}\right)  $, one finds
\begin{equation}
v\left(  r\right)  =-\sqrt{1-f\left(  r\right)  }.\label{eq:v(r)}%
\end{equation}
In the PG coordinate, one can use eqn. $\left(  \ref{eq:U-PE}\right)  $ to
show that eqn. $\left(  \ref{eq:T and X}\right)  $ becomes%
\begin{align}
T &  =-\left(  \partial_{t}I+qA_{t}\right)  -v\left(  r\right)  \left(
\partial_{r}I+qA_{r}\right)  ,\nonumber\\
X^{2} &  =\left(  \partial_{r}I+qA_{r}\right)  ^{2}+\frac{h^{ab}\left(
x\right)  \left(  \partial_{a}I+qA_{a}\right)  \left(  \partial_{b}%
I+qA_{b}\right)  }{C\left(  r^{2}\right)  }.\label{eq:TandX}%
\end{align}
Note that $\partial_{t}$ and $\partial_{t_{p}}$ are Killing vectors in the
Schwarzschild-like coordinate and the PG coordinate, respectively. For the
energy-momentum vector $p_{\mu}$, eqn. $\left(  \ref{eq:Tp-T}\right)  $ gives
us that its Killing energies associated with $\partial_{t}$ and $\partial
_{t_{p}}$ are the same. Let $\omega$ denote the Killing energies in the PG and
Schwarzschild-like coordinates. Explicitly, one has $\omega=\partial_{t}^{\mu
}p_{\mu}=\partial_{t_{p}}^{\mu}p_{\mu}$, which is a constant.

\subsection{Massive and Charged Particle to $\mathcal{O}\left(  m_{p}%
^{-2}\right)  $}

Since $\omega=-\partial_{t_{p}}I$ is the conserved energy of the particle, we
can separate $t$ from other variables. Thus, we employ the following ansatz
for the action $I$%
\begin{equation}
I=-\omega t+W\left(  r\right)  +\Theta\left(  x\right)  . \label{eq:I}%
\end{equation}
The vector potential $A_{\mu}$ is assumed to be given by
\begin{equation}
A_{\mu}=A_{t}\left(  r\right)  \delta_{\mu t}, \label{eq:vector-oneform}%
\end{equation}
which is true for charged static black holes in most cases. Putting the ansatz
$\left(  \ref{eq:I}\right)  $ into eqn. $\left(  \ref{eq:TandX}\right)  $, we
have%
\begin{align}
T  &  =\tilde{\omega}\left(  r\right)  -v\left(  r\right)  p_{r},\nonumber\\
X^{2}  &  =p_{r}^{2}+\frac{h^{ab}\left(  x\right)  \partial_{a}\Theta\left(
x\right)  \partial_{b}\Theta\left(  x\right)  }{C\left(  r^{2}\right)  },
\end{align}
where $\tilde{\omega}\left(  r\right)  =\omega-qA_{t}\left(  r\right)  $ and
$p_{r}=\partial_{r}W$. The method of separation of variables gives the
differential equation for $\Theta\left(  x\right)  $%
\begin{equation}
h^{ab}\left(  x\right)  \partial_{a}\Theta\left(  x\right)  \partial_{b}%
\Theta\left(  x\right)  =\lambda, \label{eq:HJh}%
\end{equation}
where is $\lambda$ is a constant and determined by $h^{ab}\left(  x\right)  $.
Thus, one has%
\begin{equation}
X^{2}=p_{r}^{2}+\frac{\lambda}{C\left(  r^{2}\right)  },
\end{equation}
and eqn. $\left(  \ref{eq:Deformed-HJ}\right)  $ becomes an ordinary
differential equation for $W\left(  r\right)  $.

We consider a particle with mass $m$ and charge $q$. Solving eqn.$\left(
\ref{eq:Deformed-HJ}\right)  $ for $p_{r}$ gives
\begin{equation}
p_{r}^{\pm}=\frac{\pm\sqrt{\Lambda}+\sqrt{1-f}}{f}\tilde{\omega}\left(
r\right)  \mp\frac{C_{1}}{m_{p}^{2}}\frac{\left[  1\pm\sqrt{\left(
1-f\right)  \Lambda}\right]  ^{4}\tilde{\omega}^{3}\left(  r\right)  }%
{f^{4}\sqrt{\Lambda}}+\mathcal{O}\left(  m_{p}^{-4}\right)  , \label{eq:pr-4D}%
\end{equation}
where +/$-$ denotes the outgoing/ingoing solutions and $\Lambda=1-\frac
{m^{2}+\frac{\lambda}{C\left(  r^{2}\right)  }}{\tilde{\omega}^{2}\left(
r\right)  }$. Here, $p_{r}^{+}$ has a pole at $r=r_{h}$. To obtain the residue
of $p_{r}^{+}$ at $r=r_{h}$, one expands $f\left(  r\right)  $ and $C\left(
r^{2}\right)  $ at $r=r_{h}$
\begin{align}
f\left(  r\right)   &  \sim2\kappa\left(  r-r_{h}\right)  \left[  1+\eta
\kappa\left(  r-r_{h}\right)  +\theta\kappa^{2}\left(  r-r_{h}\right)
^{2}+\rho\kappa^{3}\left(  r-r_{h}\right)  ^{3}\right]  ,\nonumber\\
C\left(  r^{2}\right)   &  \sim C\left(  r_{h}^{2}\right)  \left[
1+c_{1}\kappa\left(  r-r_{h}\right)  +c_{2}\kappa^{2}\left(  r-r_{h}\right)
^{2}+c_{3}\kappa^{3}\left(  r-r_{h}\right)  ^{3}\right]  ,\nonumber\\
\tilde{\omega}\left(  r\right)   &  \sim\tilde{\omega}\left(  r_{h}\right)
\left[  1+\omega_{1}\kappa\left(  r-r_{h}\right)  +\omega_{2}\kappa^{2}\left(
r-r_{h}\right)  ^{2}+\omega_{3}\kappa^{3}\left(  r-r_{h}\right)  ^{3}\right]
\label{eq:fandC-Expansion}%
\end{align}
where $\kappa=f^{\prime}\left(  r_{h}\right)  /2$. Using the residue theory
for the semi circles, we get%
\begin{equation}
\operatorname{Im}W_{+}\left(  r\right)  =\frac{\tilde{\omega}\left(
r_{h}\right)  \pi}{\kappa}\left(  1+\Delta\right)  , \label{eq:ImW}%
\end{equation}
where we define
\begin{align}
\Delta &  =\frac{C_{1}}{2m_{p}^{2}}\left[  \frac{\delta_{\lambda}\lambda
}{C\left(  r_{h}^{2}\right)  }+\delta_{m}m^{2}+\delta_{\omega}\tilde{\omega
}^{2}\left(  r_{h}\right)  \right]  +\mathcal{O}\left(  m_{p}^{-4}\right)
,\nonumber\\
\delta_{\lambda}  &  =3+2c_{1}^{2}-2c_{2}+12\eta+12\eta^{2}-6\theta
+c_{1}\left(  6+6\eta-2\omega_{1}\right)  -6\omega_{1}-6\eta\omega_{1}%
+2\omega_{2},\nonumber\\
\delta_{m}  &  =-1+12\eta^{2}-6\theta+\eta\left(  4-6\omega_{1}\right)
-2\omega_{1}+2\omega_{2},\nonumber\\
\delta_{\omega}  &  =40\eta^{3}-12\theta+8\rho+\eta^{2}\left(  24-60\omega
_{1}\right)  -3\omega_{1}+24\theta\omega_{1}+12\omega_{1}^{2}\nonumber\\
&  -2\omega_{1}^{2}+12\omega_{2}-12\omega_{1}\omega_{2}+\eta\left(
2-40\theta-36\omega_{1}+24\omega_{1}^{2}+24\omega_{2}\right)  -6\omega_{3}.
\label{eq:Delta-ff}%
\end{align}
On the other hand, one can use eqns. $\left(  \ref{eq:fandC-Expansion}\right)
$ to expand $p_{r}^{-}$ at $r=r_{h}$. It turns out that the residue of
$p_{r}^{-}$ at $r=r_{h}$ is zero. Hence, we have $\operatorname{Im}%
W_{-}\left(  r\right)  =0$. As shown in \cite{SF}, the probability of a
particle tunneling from inside to outside the horizon is%
\begin{equation}
P_{emit}\propto\exp\left[  -\frac{2}{\hbar}\left(  \operatorname{Im}%
W_{+}-\operatorname{Im}W_{-}\right)  \right]  . \label{eq:Prob-Emission}%
\end{equation}
There is a Boltzmann factor in $P_{emit}$ with an effective temperature, which
is
\begin{equation}
T_{eff}=\frac{T_{0}}{1+\Delta}. \label{eq:TeffMassive}%
\end{equation}
It is interesting to note that we have calculated $\Delta$ in the static
preferred frame in \cite{SF}. For emitted particles with mass $m$ and charge
$q,$ we found%
\begin{equation}
\Delta=-\frac{C_{1}}{2m_{p}^{2}}\left[  2\left(  3\omega_{1}-\eta\right)
\tilde{\omega}^{2}\left(  r_{h}\right)  +\frac{\lambda}{C\left(  r_{h}%
^{2}\right)  }+m^{2}\right]  +\mathcal{O}\left(  m_{p}^{-4}\right)  .
\label{eq:Delta-Static}%
\end{equation}

\subsection{Massless and Neutral Particle to All Orders}

We now work with a particle with $m=0$ and $q=0$. To get all order result, we
let $\lambda=0$. Note that one has $\lambda=0$ for the solution in a 2D black
hole or the s-wave solution in a 4D spherically symmetric black hole. In this
case, we can use the following ansatz for the action $I$
\begin{equation}
I=-\omega t+W\left(  r\right)  .
\end{equation}
Hence, the deformed Hamilton-Jacobi equation becomes%
\begin{equation}
\left(  \omega-vp_{r}\right)  ^{2}=p_{r}^{2}\left(
%TCIMACRO{\dsum \limits_{n=0}}%
%BeginExpansion
{\displaystyle\sum\limits_{n=0}}
%EndExpansion
C_{n}\frac{p_{r}^{2n}}{m_{p}^{2n}}\right)  ^{2},\label{eq:Deformed-HJ-2D}%
\end{equation}
where $p_{r}=\partial_{r}W$. We will prove by induction that the solutions to
eqn. $\left(  \ref{eq:Deformed-HJ-2D}\right)  $ take the form of%
\begin{equation}
p_{r}^{\pm}=\frac{\pm\omega}{1\pm v}\left[  1+\sum_{i=1}^{\infty}\frac
{\omega^{2i}}{m_{p}^{2i}}\sum\limits_{j=0}^{i-1}\frac{C_{i,j}}{\left(  1\pm
v\right)  ^{3i-j}}\right]  ,\label{eq:Pr-Plus-Minus}%
\end{equation}
where $C_{i,j}$ is determined by $C_{n}$. In fact, it is easy to see that%
\begin{equation}
p_{r}^{\pm}=\frac{\pm\omega}{1\pm v}\left[  1-\frac{C_{1}\omega^{2}}{m_{p}%
^{2}\left(  1\pm v\right)  ^{3}}+\frac{\omega^{4}\left[  3C_{1}^{2}%
-C_{2}\left(  1\pm v\right)  \right]  }{m_{p}^{4}\left(  1\pm v\right)  ^{6}%
}+\mathcal{O}\left(  \frac{\omega^{6}}{m_{p}^{6}}\right)  \right]  .
\end{equation}
Suppose for some integer $N>$ $1$ such that
\begin{equation}
p_{r,N}^{\pm}=\frac{\pm\omega}{1\pm v}\left[  1+\sum_{i=1}^{N}\frac
{\omega^{2i}}{m_{p}^{2i}}\sum\limits_{j=0}^{i-1}\frac{C_{i,j}}{\left(  1\pm
v\right)  ^{3i-j}}\right]
\end{equation}
satisfy eqn. $\left(  \ref{eq:Deformed-HJ-2D}\right)  $ at $\mathcal{O}\left(
\frac{\omega^{2N}}{m_{p}^{2N}}\right)  $. Plugging $p_{r}=p_{r,N}^{\pm}%
+\frac{\alpha^{\pm}\omega^{2N+2}}{m_{p}^{2N+2}}\frac{\omega}{1\pm v}$ into
eqn. $\left(  \ref{eq:Deformed-HJ-2D}\right)  ,$ one finds that the terms of
$\mathcal{O}\left(  \frac{\omega^{2N+2}}{m_{p}^{2N+2}}\right)  $ gives%
\begin{equation}
\alpha^{\pm}=\frac{P_{N-1}\left(  1\pm v\right)  }{\left(  1\pm v\right)
^{3\left(  N+1\right)  }},
\end{equation}
where $P_{N-1}\left(  x\right)  $ is some polynomial of $x$ with degree $N-1$.
This completes the proof that eqn. $\left(  \ref{eq:Pr-Plus-Minus}\right)  $
are solutions to eqn. $\left(  \ref{eq:Deformed-HJ-2D}\right)  $ to all
orders. We define the residue of $\frac{1}{\left(  1+v\right)  ^{n}}$ at
$r=r_{h}$ as
\begin{equation}
\text{Res}\left(  \frac{1}{\left(  1+v\right)  ^{n}},r_{h}\right)
=\frac{R_{n}}{\kappa},
\end{equation}
where $R_{1}=1$. Thus, one obtains%
\begin{equation}
\operatorname{Im}W^{+}=\frac{\omega\pi}{\kappa}\left(  1+\Delta\right)
,\text{ }\operatorname{Im}W^{-}=0,\label{eq:IMW-massless}%
\end{equation}
where we define%
\begin{equation}
\Delta=\sum_{i=1}^{\infty}\frac{\delta_{i}\omega^{2i}}{m_{p}^{2i}}\text{ with
}\delta_{i}=\left(  \sum\limits_{j=0}^{i-1}C_{i,j}R_{3i+1-j}\right)  .
\end{equation}
Using eqn. $\left(  \ref{eq:Prob-Emission}\right)  $ gives the effective
temperature%
\begin{equation}
T_{eff}=\frac{T_{0}}{1+\Delta}.\label{eq:TeffMassless}%
\end{equation}

\subsection{Discussion}

When we use the residue theory for the semi circles to give eqns. $\left(
\ref{eq:ImW}\right)  $ and $\left(  \ref{eq:IMW-massless}\right)  $, an
assumption proposed in \cite{SF} is needed. The assumption requires that the
singularity structure of $\partial_{r}I$ except the order of the pole at
$r=r_{h}$ do not change after the MDR is introduced. It follows that
$\omega\lesssim m_{p}$. A complete theory of quantum gravity might been needed
to justify this assumption.

In most works of the dispersive models, much attention have been paid to the
modifications of the asymptotic spectrum. Since the tunneling across the
horizon takes place near the horizon, the near horizon spectrum of radiations
is computed in the paper. The effects of scattering off the background need to
be included if the asymptotic spectrum at infinity is considered. In most
works, 2D spacetime has been considered. The higher order terms in the MDR
violate conformal invariance of 2D spacetime, hence there is some scattering.
Our calculations show that the spectrum of radiation near the horizon is close
to a perfect thermal spectrum in the dispersive models. The thermal asymptotic
spectrum has been recovered at leading order in previous studies. Thus, the
energy fluxes of radiations are not significantly affected by the effects of
scattering off the background. However, as noted in \cite{SF}, the scattering
effects might dramatically change the spectrum of radiations in the dispersive
models with the static preferred frame.

\section{Entropy in Brick Wall Model}

\label{Sec:EBWM}

Bekenstein and Hawking showed that the entropy of a black hole is proportional
to the area of the
horizon\cite{EBWM-Bekenstein:1972tm,EBWM-Bekenstein:1973ur,EBWM-Hawking:1976de}%
. Although all the evidences suggest that the Bekenstein-Hawking entropy is
the thermodynamic entropy, the statistical origin of the black hole entropy
has not yet been fully understood. It appears that an unavoidable candidate
for the statistical origin is the entropy of the thermal atmosphere of the
black hole.

However, the entropy diverges when we attempt to calculate the entropy of the
thermal atmosphere. There are two kinds of divergences. The first one is due
to infinite volume of the system, which has to do with the contribution from
the vacuum surrounding the system at large distances and is of little
relevance here. The second one arises from the infinite volume of the deep
throat region near the horizon. To regulate the divergences, t' Hooft
\cite{EBWM-'tHooft:1984re} proposed the brick wall model for a scalar field
$\phi$, where two brick wall cutoffs are introduced at some small distance
$r_{\varepsilon}$ from the horizon and at a large distance $L\gg r_{h}$,
\begin{equation}
\phi=0\text{ \ at }r=r_{h}+r_{\varepsilon}\text{ and }r=L\text{.}
\label{eq:DirichletBoundary}%
\end{equation}

In the following, we will use the brick wall model to calculate the entropy of
a scalar field for a 4D spherically symmetric black hole and a 2D one. For a
4D spherically symmetric black hole, the entropy will be calculated to
$O\left(  m_{p}^{-2}\right)  $. For the 2D black hole, we will obtain all
order results in the cases with the static and free-fall preferred frames for
comparison. For simplicity, we assume that the scalar field is massless and neutral.

\subsection{4D Spherically Symmetric Black Hole}

For a 4D spherically symmetric black hole with the Schwarzschild-like
coordinate%
\begin{equation}
ds^{2}=f\left(  r\right)  dt^{2}-\frac{dr^{2}}{f\left(  r\right)  }-C\left(
r^{2}\right)  \left(  d\theta^{2}+\sin^{2}\theta d\phi^{2}\right)  ,
\end{equation}
we have shown that $\lambda=\left(  l+\frac{1}{2}\right)  ^{2}\hbar^{2}$ with
the angular momentum $l=0,1,\cdots$ and the corresponding degeneracy is
$2l+1$\cite{SF}. Thus, the atmosphere entropy of a massless scalar field can
be expressed in the form of%
\begin{equation}
S=\int\left(  2l+1\right)  dl\int d\omega\frac{dn\left(  \omega,l\right)
}{d\omega}s_{\omega,l},
\end{equation}
where $\omega$ is the Killing energy associated with $t$, $l$ is the angular
momentum, $n\left(  \omega,l\right)  $ is the number of one-particle states
not exceeding $\omega$ with fixed value of angular momentum $l$, and
$s_{\omega,l}$ is the thermal entropy per mode. Taking the MDR corrections to
both $n\left(  \omega,l\right)  $ and the Hawking temperature into
consideration, we used the brick wall model to calculate this entropy to all
orders in \cite{SF}, where the static preferred frame is used to import a MDR
to the black hole background. By contrast, here the MDR corrected atmosphere
entropy of the black hole is computed to $\mathcal{O}\left(  m_{p}%
^{-2}\right)  $ in the free-fall scenario.

For particles emitted in a wave mode labelled by energy $\omega$ and $l$, we
find that%
\begin{align*}
&  \left(  \text{Probability for a black hole to emit a particle in this
mode}\right) \\
&  =\exp\left(  -\frac{\omega}{T_{eff}}\right)  \times(\text{Probability for a
black hole to absorb a particle in the same mode}),
\end{align*}
where $T_{eff}$ is given by eqn. $\left(  \ref{eq:TeffMassive}\right)  $. The
above relation for the usual dispersion relation was obtained by Hartle and
Hawking \cite{EBWM-Hartle:1976tp} using semiclassical analysis. Neglecting
back-reaction, detailed balance condition requires that the ratio of the
probability of having $N$ particles in a particular mode to the probability of
having $N-1$ particles in the same mode is $\exp\left(  -\frac{\omega}%
{T_{eff}}\right)  .$ One then follows the argument in \cite{SF} to get the
average number $n_{\omega,l}$ in the mode%
\begin{equation}
n_{\omega,l}=n\left(  \frac{\omega}{T_{eff}}\right)  , \label{eq:Number}%
\end{equation}
where we define%
\begin{equation}
n\left(  x\right)  =\frac{1}{\exp x-\left(  -1\right)  ^{\epsilon}}.
\end{equation}
Note that $\epsilon=0$ for bosons and $\epsilon=1$ for fermions. The von
Neumann entropy for the mode is%
\begin{equation}
s_{\omega,l}=\left[  n_{\omega,l}+\left(  -1\right)  ^{\epsilon}\right]
\ln\left[  1+\left(  -1\right)  ^{\epsilon}n_{\omega,l}\right]  -n_{\omega
,l}\ln n_{\omega,l}.
\end{equation}
Moreover, the entropy per mode $s_{\omega,l}$ can be put in the form of%
\begin{equation}
s_{\omega,l}=s\left(  \frac{\omega}{T_{eff}}\right)  ,
\end{equation}
where the $s\left(  x\right)  $ is given by%
\begin{equation}
s\left(  x\right)  =\frac{\left(  -1\right)  ^{\epsilon}\exp x}{\exp x-\left(
-1\right)  ^{\epsilon}}\ln\left[  \frac{\exp x}{\exp x-\left(  -1\right)
^{\epsilon}}\right]  +\frac{\ln\left[  \exp x-\left(  -1\right)  ^{\epsilon
}\right]  }{\exp x-\left(  -1\right)  ^{\epsilon}}. \label{eq:S(x)}%
\end{equation}
Defining $u=\frac{\omega}{T_{0}}$ and expanding $s_{\omega,l}$ to
$\mathcal{O}\left(  m_{p}^{-2}\right)  $ give%
\begin{equation}
s_{\omega,l}\approx s\left(  u\right)  +s^{\prime}\left(  u\right)  u\Delta,
\label{eq:Entropy-4D-Expansion}%
\end{equation}
where one has
\begin{equation}
\Delta=\frac{C_{1}\delta_{\lambda}z}{2}+\frac{C_{1}\delta_{\omega}T_{0}^{2}%
}{2m_{p}^{2}}u^{2}.
\end{equation}

In the brick wall model, t' Hooft' found that the number of one-particle
states not exceeding $\omega$ fixed value $l$ is%
\begin{equation}
n\left(  \omega,l\right)  =\frac{1}{2\pi\hbar}%
%TCIMACRO{\doint }%
%BeginExpansion
{\displaystyle\oint}
%EndExpansion
p_{r}dr, \label{eq:Number-States}%
\end{equation}
where the integral $%
%TCIMACRO{\doint }%
%BeginExpansion
{\displaystyle\oint}
%EndExpansion
p_{r}dr$ was calculated in the Schwarzschild-like coordinate. Nevertheless, we
calculate $p_{r}^{\pm}$ in the PG coordinate in section \ref{Sec:DHJM}. In
\cite{EBWM-Chowdhury:2006sk,EBWM-Akhmedov:2008ru}, the integral $%
%TCIMACRO{\doint }%
%BeginExpansion
{\displaystyle\oint}
%EndExpansion
p_{r}dr$ has been found invariant under canonical transformations. Hence, the
number states $n\left(  \omega,l\right)  $ given in eqn. $\left(
\ref{eq:Number-States}\right)  $ is the same in the Schwarzschild-like and PG
coordinates and one does not need to re-calculate it in different coordinates.
Define the radial wave number $k\left(  r,l,\omega\right)  $ by%
\begin{equation}
k^{\pm}\left(  r,l,\omega\right)  =p_{r}^{\pm}, \label{eq:k(r)}%
\end{equation}
as long as $p_{r}^{\pm2}\geq0$, and $k^{\pm}\left(  r,l,\omega\right)  =0$
otherwise. With these two Dirichlet boundaries, one finds that the number of
one-particle states not exceeding $\omega$ fixed value $l$ is%
\begin{equation}
n\left(  \omega,l\right)  =\frac{1}{2\pi\hbar}\left[  \int_{r_{h}%
+r_{\varepsilon}}^{L}k^{+}\left(  r,l,\omega\right)  dr+\int_{L}%
^{r_{h}+r_{\varepsilon}}k^{-}\left(  r,l,\omega\right)  dr\right]  .
\end{equation}
The $p_{r}^{\pm}$ in eqn. $\left(  \ref{eq:k(r)}\right)  $ are given by eqn.
$\left(  \ref{eq:pr-4D}\right)  $ in section \ref{Sec:DHJM}. For a massless
and neutral scalar field, one thus has%
\begin{equation}
p_{r}^{\pm}=\frac{\pm\sqrt{\Lambda}+\sqrt{1-f}}{f}\omega\mp\frac{C_{1}}%
{m_{p}^{2}}\frac{\left[  1\pm\sqrt{\Lambda\left(  1-f\right)  }\right]
^{4}\omega^{3}}{f^{4}\sqrt{\Lambda}}+\mathcal{O}\left(  m_{p}^{-4}\right)  ,
\label{eq:Pr-4D}%
\end{equation}
where $\Lambda=1-f\frac{\left(  l+\frac{1}{2}\right)  ^{2}\hbar^{2}}{C\left(
r^{2}\right)  \omega^{2}}$.

Integrating by parts, one finds the entropy becomes
\begin{equation}
S=-\frac{1}{\pi\hbar}\int\left(  2l+1\right)  dl\int du\frac{\partial
s_{\omega,l}}{\partial u}\int_{r_{h}+r_{\varepsilon}}^{L}k\left(
r,l,\omega\right)  dr, \label{eq:Entropy-4D}%
\end{equation}
where we define $k\left(  r,l,\omega\right)  =\frac{k^{+}\left(
r,l,\omega\right)  -k^{-}\left(  r,l,\omega\right)  }{2}$. Plugging eqns.
$\left(  \ref{eq:Pr-4D}\right)  $ and $\left(  \ref{eq:Entropy-4D-Expansion}%
\right)  $ into eqn. $\left(  \ref{eq:Entropy-4D}\right)  $ and performing the
$l$ integral which runs over the region where $\Lambda>0$, we find that the
entropy to $\mathcal{O}\left(  m_{p}^{-2}\right)  $ becomes%
\begin{align}
S  &  \approx\frac{C\left(  r_{h}^{2}\right)  T_{0}^{3}}{\pi\hbar^{3}}\int
u^{2}du\int_{r_{h}+r_{\varepsilon}}^{L}\frac{C\left(  r^{2}\right)  }{C\left(
r_{h}^{2}\right)  }\frac{dr}{f^{2}}\left\{  2s\left(  u\right)  +\frac
{C_{1}\delta_{\omega}T_{0}^{2}}{m_{p}^{2}}u^{3}s^{\prime}\left(  u\right)
+\frac{2C_{1}\delta_{\lambda}C\left(  r^{2}\right)  T_{0}^{2}}{3fC\left(
r_{h}^{2}\right)  m_{p}^{2}}u^{3}s^{\prime}\left(  u\right)  \right.
\nonumber\\
&  \left.  -\frac{5C_{1}T_{0}^{2}u^{2}s\left(  u\right)  }{2m_{p}^{2}f^{3}%
}\left(  12-\frac{48f}{5}+\frac{f^{2}}{2}-\frac{7f^{4}}{32}\right)  \right\}
,
\end{align}
where second and third terms in the bracket come from the MDR corrections to
the Hawking temperature and the fourth term from the MDR corrections to
$n\left(  \omega,l\right)  $. Focusing on the divergent parts near horizon, we
obtain for the nonnegative integers $a$ and $n$%
\begin{equation}
\int_{r_{h}+r_{\varepsilon}}\frac{C^{a}\left(  r^{2}\right)  }{C^{a}\left(
r_{h}^{2}\right)  }\frac{dr}{f\left(  r\right)  ^{n+1}}\sim\frac{1}%
{2^{n+1}\kappa}\sum_{k=0}^{n-1}\frac{\tilde{f}_{k}^{n,a}}{n-k}\left(  \kappa
r_{\varepsilon}\right)  ^{k-n}-\frac{\tilde{f}_{n}^{n,a}}{2^{n+1}\kappa}%
\ln\kappa r_{\varepsilon}, \label{eq:CFIntegrals}%
\end{equation}
where we expand $f\left(  r\right)  ^{-n}$ and $\frac{C^{a}\left(
r^{2}\right)  }{C^{a}\left(  r_{h}^{2}\right)  }$ at $r=r_{h}$%
\begin{align}
f\left(  r\right)  ^{-n}  &  =2^{-n}\kappa^{-n}\left(  r-r_{h}\right)  ^{-n}%
%TCIMACRO{\dsum \limits_{j=0}^{\infty}}%
%BeginExpansion
{\displaystyle\sum\limits_{j=0}^{\infty}}
%EndExpansion
f_{j}^{n}\kappa^{j}\left(  r-r_{h}\right)  ^{j},\nonumber\\
\frac{C^{a}\left(  r^{2}\right)  }{C^{a}\left(  r_{h}^{2}\right)  }  &
=\sum_{i=0}c_{i}^{a}\kappa^{i}\left(  r-r_{h}\right)  ^{i}, \label{eq:f and c}%
\end{align}
and define $\tilde{f}_{k}^{n,a}=\sum_{j=0}^{k}f_{j}^{n+1}c_{k-j}^{a}$. In eqn.
$\left(  \ref{eq:CFIntegrals}\right)  ,\,$\ we neglect finite terms as $\kappa
r_{\varepsilon}\rightarrow0$ and terms involving $L$. Note that we define
$\kappa=\frac{f^{\prime}\left(  r_{h}\right)  }{2}$ which is the surface
gravity for the black hole and hence $T_{0}=\frac{\hbar\kappa}{2\pi}$. Thus,
the divergent part of entropy near the horizon to $\mathcal{O}\left(
m_{p}^{-2}\right)  $ is%
\begin{align}
S  &  \sim\frac{C\left(  r_{h}^{2}\right)  \kappa^{2}}{16\pi^{4}}\int s\left(
u\right)  u^{2}du\left(  \frac{1}{\kappa r_{\varepsilon}}-\left(  c_{1}%
-2\eta\right)  \ln\kappa r_{\varepsilon}\right) \nonumber\\
&  -\frac{C\left(  r_{h}^{2}\right)  \kappa^{2}}{16\pi^{4}}\frac{m_{p}%
^{2}\kappa^{2}}{4\pi^{2}}\int s\left(  u\right)  u^{4}du\left\{  \frac
{5C_{1}\delta_{\omega}}{2}\left(  \frac{\tilde{f}_{0}^{1,1}}{\kappa
r_{\varepsilon}}-\tilde{f}_{1}^{1,1}\ln\kappa r_{\varepsilon}\right)  \right.
\nonumber\\
&  +\frac{5C_{1}\delta_{\lambda}}{6}\left[  \sum_{k=0}^{1}\frac{\tilde{f}%
_{k}^{2,2}}{2-k}\left(  \kappa r_{\varepsilon}\right)  ^{k-2}-\tilde{f}%
_{2}^{2,2}\ln\kappa r_{\varepsilon}\right] \nonumber\\
&  +5C_{1}\left[  \frac{12}{32}\sum_{k=0}^{3}\frac{\tilde{f}_{k}^{4,1}}%
{4-k}\left(  \kappa r_{\varepsilon}\right)  ^{k-4}-\frac{3}{5}\sum_{k=0}%
^{2}\frac{\tilde{f}_{k}^{3,1}}{3-k}\left(  \kappa r_{\varepsilon}\right)
^{k-3}+\frac{\tilde{f}_{0}^{1,1}}{8\kappa r_{\varepsilon}}\right. \nonumber\\
&  \left.  \left.  +\left(  \frac{7\tilde{f}_{0}^{0,1}}{64}-\frac{\tilde
{f}_{1}^{1,1}}{8}+\frac{3\tilde{f}_{3}^{3,1}}{5}-\frac{12\tilde{f}_{4}^{4,1}%
}{32}\right)  \ln\kappa r_{\varepsilon}\right]  \right\}  .
\end{align}

\subsection{2D Black Hole in Free-fall Scenario}

Consider a 2D black hole with the metric of
\begin{equation}
ds^{2}=f\left(  r\right)  dt^{2}-\frac{dr^{2}}{f\left(  r\right)  },
\label{eq:2D-BH}%
\end{equation}
in the Schwarzschild-like coordinate. The atmosphere entropy of a massless
scalar field is%
\begin{equation}
S=\int d\omega\frac{dn\left(  \omega,l\right)  }{d\omega}s_{\omega},
\end{equation}
where $s_{\omega}$ is the entropy per mode. Define the radial wave number
$k\left(  r,\omega\right)  $ by%
\begin{equation}
k^{\pm}\left(  r,\omega\right)  =p_{r}^{\pm}, \label{eq:k(r)-2D}%
\end{equation}
as long as $p_{r}^{\pm2}\geq0$, and $k^{\pm}\left(  r,\omega\right)  =0$
otherwise. The $p_{r}^{\pm}$ in eqn. $\left(  \ref{eq:k(r)-2D}\right)  $ are
given in eqn. $\left(  \ref{eq:Pr-Plus-Minus}\right)  $ in section
\ref{Sec:DHJM}. The number of one-particle states not exceeding $\omega$ is%
\begin{equation}
n\left(  \omega\right)  =\frac{1}{2\pi\hbar}\int_{r_{h}+r_{\varepsilon}}%
^{L}\left(  k_{r}^{+}-k_{r}^{-}\right)  dr.
\end{equation}
Defining the coefficients $\sigma_{s}^{q}$ by%
\begin{equation}
\Delta^{q}=\left(  \frac{m_{p}\kappa}{2\pi}\right)  ^{2q}u^{2q}\sum
_{s=0}^{\infty}\sigma_{s}^{q}\left(  \frac{m_{p}\kappa}{2\pi}\right)
^{2s}u^{2s},
\end{equation}
one has for $s_{\omega}$%
\begin{align}
s_{\omega}  &  =s\left(  \frac{\omega}{T_{eff}}\right) \nonumber\\
&  =%
%TCIMACRO{\dsum \limits_{r=0}^{\infty}}%
%BeginExpansion
{\displaystyle\sum\limits_{r=0}^{\infty}}
%EndExpansion
\sum_{q=0}^{r}\frac{s^{\left(  q\right)  }\left(  u\right)  u^{q+2r}%
\sigma_{r-q}^{q}}{q!}\left(  \frac{m_{p}\kappa}{2\pi}\right)  ^{2r},
\end{align}
where $u=\frac{\omega}{T_{0}}$ and we use eqn. $\left(  \ref{eq:TeffMassless}%
\right)  $ for $T_{eff}$. Expanding $\left(  1+v\right)  ^{i}$ at $r=r_{h}$
\begin{equation}
\left(  1+v\right)  ^{i}=\kappa^{i}\left(  r-r_{h}\right)  ^{i}\sum
_{j=0}^{\infty}v_{j}^{i}\kappa^{j}\left(  r-r_{h}\right)  ^{j},
\end{equation}
we find the divergent part of entropy near the horizon becomes%
\begin{align}
S  &  \sim-\frac{1}{4\pi^{2}}\ln\kappa r_{\varepsilon}%
%TCIMACRO{\dsum \limits_{r=0}^{\infty}}%
%BeginExpansion
{\displaystyle\sum\limits_{r=0}^{\infty}}
%EndExpansion
\sum_{i=0}^{\infty}\sum_{q=0}^{r}\frac{\left(  2i+1\right)  \eta_{i}%
\sigma_{r-q}^{q}}{q!}\left(  \frac{m_{p}\kappa}{2\pi}\right)  ^{2i+2r}\int
dus^{\left(  q\right)  }\left(  u\right)  u^{q+2r+2i}\nonumber\\
&  +\frac{1}{4\pi^{2}}\sum_{l=1}^{\infty}\frac{1}{l\left(  \kappa
r_{\varepsilon}\right)  ^{l}}\left(  \frac{m_{p}\kappa}{2\pi}\right)
^{2\left\lceil l/3\right\rceil }%
%TCIMACRO{\dsum \limits_{r=0}^{\infty}}%
%BeginExpansion
{\displaystyle\sum\limits_{r=0}^{\infty}}
%EndExpansion
\sum\limits_{i=0}^{\infty}\sum_{q=0}^{r}\frac{\xi_{i,l}\sigma_{r-q}^{q}}%
{q!}\left(  \frac{m_{p}\kappa}{2\pi}\right)  ^{2i+2r}\int dus^{\left(
q\right)  }\left(  u\right)  u^{q+2r+2i+2\left\lceil l/3\right\rceil },
\label{eq:entropy-2d-ff}%
\end{align}
where $\left\lceil x\right\rceil $ is the smallest integer that is not less
than$\ x$ and we define%
\begin{align}
\eta_{i>0}  &  =\sum\limits_{j=0}^{i-1}C_{i,j}v_{3i-j}^{-3i+j-1}\text{ and
}\eta_{0}=1,\nonumber\\
\xi_{i,l}  &  =\sum_{j=0}^{3i+3\left[  l/3\right]  ^{+}-l}\left(  2i+2\left[
l/3\right]  ^{+}+1\right)  C_{i+\left[  l/3\right]  ^{+},j}v_{3i-j-l+3\left[
l/3\right]  ^{+}}^{-l}.
\end{align}

\subsection{2D Black Hole in Static Scenario}

Following the conventions adopted in \cite{SF}, the MDR for a massless scalar
particle considered here takes the form of
\begin{equation}
p^{2}=\tilde{F}^{2}\left(  E\right)  ,
\end{equation}
where we define
\begin{equation}
\tilde{F}\left(  E\right)  =E\sum_{n=0}^{\infty}\frac{\tilde{C}_{n}E^{2n}%
}{m_{p}^{2n}}, \label{eq:MDR-SF}%
\end{equation}
with $\tilde{C}_{0}=1$. For a particle with the energy-momentum vector
$p_{\mu}$, the energy $\omega$ and the norm of the momentum $p$ of the
particle measured by the static observers hovering above the 2D black hole
with the metric $\left(  \ref{eq:2D-BH}\right)  $ are
\begin{align}
E  &  =\frac{p_{t}}{\sqrt{f\left(  r\right)  }},\nonumber\\
p^{2}  &  =f\left(  r\right)  p_{r}^{2}. \label{eq:momentum-BH}%
\end{align}
Relating $p_{\mu}$ to the action $I$ by $p_{\mu}=-\partial_{\mu}I$ gives the
deformed Hamilton-Jacobi equation%
\begin{equation}
X^{2}=\tilde{F}^{2}\left(  T\right)  , \label{eq:HJD-Static}%
\end{equation}
where%
\begin{equation}
T=\frac{\omega}{\sqrt{f\left(  r\right)  }},\text{ }X=\sqrt{f\left(  r\right)
}p_{r},
\end{equation}
and $\omega=\partial_{t}^{\mu}p_{\mu}$ is the Killing energy with respect to
$t$. Solving eqn. $\left(  \ref{eq:HJD-Static}\right)  $ for $p_{r},$ one
could define the radial wave number $k\left(  r,\omega\right)  $ by%
\begin{equation}
k\left(  r,\omega\right)  =\left\vert p_{r}\right\vert ,
\end{equation}
as long as $p_{r}^{2}\geq0$, and $k\left(  r,\omega\right)  =0$ otherwise. The
number of one-particle states not exceeding $\omega$ is%
\begin{equation}
n\left(  \omega\right)  =\frac{1}{\pi\hbar}\int_{r_{h}+r_{\varepsilon}}%
^{L}k_{r}dr.
\end{equation}
The entropy per mode $s_{\omega}$ is
\begin{equation}
s_{\omega}=s\left(  \frac{\omega}{T_{eff}}\right)  ,
\end{equation}
where $s\left(  x\right)  $ is given in eqn. $\left(  \ref{eq:S(x)}\right)  $
and $T_{eff}$ is the effective Hawking temperature. We calculated $T_{eff}$ in
\cite{SF} and it was given by%
\begin{equation}
T_{eff}=\frac{T_{0}}{1+\Delta},
\end{equation}
where one has
\begin{equation}
\Delta=%
%TCIMACRO{\dsum \limits_{k=1}^{\infty}}%
%BeginExpansion
{\displaystyle\sum\limits_{k=1}^{\infty}}
%EndExpansion
\eta_{0}^{2k}\zeta_{k}^{0}\frac{\omega^{2k}}{m_{p}^{2k}},
\end{equation}
and $\eta_{0}^{2k}$ and $\zeta_{k}^{0}$ are defined in \cite{SF}. Defining the
coefficients $\tilde{\sigma}_{s}^{q}$ by
\begin{equation}
\tilde{\Delta}^{q}=\left(  \frac{m_{p}\kappa}{2\pi}\right)  ^{2q}u^{2q}%
\sum_{s=0}^{\infty}\tilde{\sigma}_{s}^{q}\left(  \frac{m_{p}\kappa}{2\pi
}\right)  ^{2s}u^{2s},
\end{equation}
we find the divergent part of entropy near the horizon becomes
\begin{align}
S  &  \sim\frac{1}{4\pi^{2}}\sum_{n=0}^{\infty}%
%TCIMACRO{\dsum \limits_{r=0}^{\infty}}%
%BeginExpansion
{\displaystyle\sum\limits_{r=0}^{\infty}}
%EndExpansion
\sum_{q=0}^{r}\left(  \frac{m_{p}\kappa}{2\pi}\right)  ^{2n+2r}\frac
{2n+1}{2^{n}}\frac{\tilde{\sigma}_{r-q}^{q}}{q!}\tilde{C}_{n}\nonumber\\
&  \left(  \sum_{l=1}^{n}\frac{\tilde{f}_{n-l}^{n,0}}{l\left(  \kappa
r_{\varepsilon}\right)  ^{l}}-\tilde{f}_{n}^{n,0}\ln r_{\varepsilon}\right)
\int u^{2n+2r+q}s^{\left(  q\right)  }\left(  u\right)  du,
\label{eq:entropy-2d-sf}%
\end{align}
where $\tilde{f}_{k}^{n,a}$ with $n\geq k\geq0$ are defined in eqn. $\left(
\ref{eq:CFIntegrals}\right)  $.

\subsection{Discussion}

In \cite{SF} and this paper, we have calculated the divergent part of the near
horizon atmosphere entropy of a massless scalar field for a 4D spherically
symmetric black hole in the static and free-fall scenarios, respectively. It
appears that the divergent part in both scenarios can be presented in the form
of a Laurent series with respect to $r_{\varepsilon}$%
\begin{equation}
S\sim\frac{s_{1}^{0}}{\kappa r_{\varepsilon}}+s_{0}^{0}\ln\kappa
r_{\varepsilon}+%
%TCIMACRO{\dsum \limits_{i=1}^{\infty}}%
%BeginExpansion
{\displaystyle\sum\limits_{i=1}^{\infty}}
%EndExpansion
\frac{T_{0}^{2i}}{m_{p}^{2i}}\left(
%TCIMACRO{\dsum \limits_{j=1}^{\delta_{i}}}%
%BeginExpansion
{\displaystyle\sum\limits_{j=1}^{\delta_{i}}}
%EndExpansion
s_{j}^{i}\left(  \kappa r_{\varepsilon}\right)  ^{-j}+s_{0}^{i}\ln\kappa
r_{\varepsilon}\right)  ,\label{eq:divergent-4D}%
\end{equation}
where $\delta_{i}=2i+1$ in the static scenario and $\delta_{i}=3i+1$ in the
free-fall scenario. Although we calculated the atmosphere entropy around the
4D black hole to $\mathcal{O}\left(  m_{p}^{-2}\right)  $ in the free-fall
scenario, the 2D black hole result suggests that eqn. $\left(
\ref{eq:divergent-4D}\right)  $ might hold to all orders in this scenario. For
$s_{1}^{0}$ and $s_{0}^{0}$ in eqn. $\left(  \ref{eq:divergent-4D}\right)  $,
we find that
\begin{align}
s_{1}^{0} &  =\frac{A\kappa^{2}}{720\pi},\nonumber\\
s_{0}^{0} &  =-\frac{C\left(  r_{h}^{2}\right)  \kappa^{2}}{180}\left(
c_{1}-2\eta\right)  ,
\end{align}
where $A=4\pi C\left(  r_{h}^{2}\right)  $ is the horizon area. In the
non-dispersive scenario $\left(  m_{p}\rightarrow\infty\right)  $, the terms
$\frac{s_{1}^{0}}{\kappa r_{\varepsilon}}$ and $s_{0}^{0}\ln\kappa
r_{\varepsilon}$ are the usual leading and subleading logarithmic divergent
terms, respectively. Note that $s_{1}^{0}$ and $s_{0}^{0}$ have already been
calculated in the non-dispersive
scenario\cite{EBWM-'tHooft:1984re,EBWM-Solodukhin:1994yz,EBWM-Solodukhin:2011gn}%
.

It seems from eqn. $\left(  \ref{eq:divergent-4D}\right)  $ that the near
horizon divergence of the atmosphere entropy gets worse for the higher order
corrections in the MDR as $\kappa r_{\varepsilon}\rightarrow0$. However, the
higher order contributions in eqn. $\left(  \ref{eq:divergent-4D}\right)  $
are always accompanied with the powers of the factor $\frac{T_{0}^{2}}%
{m_{p}^{2}}=\left(  \frac{m_{p}\kappa}{2\pi}\right)  ^{2}$. Thus, one might
hope that the higher order divergent problem would become less severe if
$r_{\varepsilon}$ somehow can be related to $m_{p}$. One way to understand the
value of $r_{\varepsilon}$ is introducing the proper length for
$r_{\varepsilon}$ as%
\begin{equation}
\varepsilon=\int_{r_{h}}^{r_{h}+r_{\varepsilon}}\sqrt{g_{rr}}%
dr.\label{eq:Proper-Distance}%
\end{equation}
The brick wall is put at $r=r_{h}+r_{\varepsilon}$ to cut off the unknown
quantum physics of gravity. In this sense, the invariant distance of the wall
from the horizon $\varepsilon$ could be given by $\varepsilon\sim m_{p}$.
Thus, we could define $\alpha$ such as $\varepsilon=\alpha m_{p}$. Indeed in
the 't Hooft's original calculation, equating $\frac{s_{1}^{0}}{\kappa
r_{\varepsilon}}$ to the Black hole's Bekenstein-Hawking entropy $S_{BH}%
=\frac{A}{4m_{p}^{2}}$ gives%
\begin{equation}
r_{\varepsilon}=\frac{\kappa m_{p}^{2}}{180\pi}.\label{eq:r-epsilon}%
\end{equation}
Note that eqn. $\left(  \ref{eq:Proper-Distance}\right)  $ depends on the
chosen coordinate system. In the scenario without the MDR, a natural choice is
that $\varepsilon$ is measured along a static time slice. Thus, eqn. $\left(
\ref{eq:Proper-Distance}\right)  $ is calculated in the Schwarzschild-like
coordinate\cite{EBWM-'tHooft:1984re}. Assuming $r_{\varepsilon}\ll r_{h}$, one
finds from eqn. $\left(  \ref{eq:Proper-Distance}\right)  $
\begin{equation}
r_{\varepsilon}\approx\frac{\kappa\varepsilon^{2}}{2}.
\end{equation}
Thus, eqn. $\left(  \ref{eq:r-epsilon}\right)  $ gives%
\begin{equation}
\alpha=\sqrt{\frac{1}{90\pi}},
\end{equation}
where we reproduce 't Hooft's result.

In the static scenario, it is still natural to assume that $\varepsilon$ is
measured along a static time slice. If we let $\varepsilon=\alpha m_{p}$ in
eqn. $\left(  \ref{eq:divergent-4D}\right)  $, we find the atmosphere entropy
around the horizon becomes%
\begin{equation}
S\sim\frac{A}{4m_{p}^{2}}\tilde{s}_{0}+2s_{0}^{0}\ln\kappa m_{p}+\text{Finite
terms as }m_{p}\kappa\rightarrow0, \label{eq:entropy-sf-4d}%
\end{equation}
where $\tilde{s}_{0}$ was given in \cite{SF}. The leading divergent
coefficient $\tilde{s}_{0}$ is determined by the coefficients $\tilde{C}_{n}$
in the MDR $\left(  \ref{eq:MDR-SF}\right)  $ and $f_{j}^{n}$ and $c_{i}^{a}$
which are defined in eqn. $\left(  \ref{eq:f and c}\right)  $. For a general
black hole, $f_{j}^{n}$ and $c_{i}^{a}$ could depend on the parameters of the
black hole. However, they are pure numbers for a Schwarzschild black hole.
Thus, for a Schwarzschild black hole, $\tilde{s}_{0}$ does not depend on the
black hole's properties and the leading divergent term in eqn. $\left(
\ref{eq:entropy-sf-4d}\right)  $ scales with the horizon area $A$.

In the free-fall scenario, one might prefer that the proper length
$\varepsilon$ is measured on a time slice orthogonal to the free-fall world
lines\cite{EBWM-Jacobson:2007jx}. In this case, eqn. $\left(
\ref{eq:Proper-Distance}\right)  $ for $\varepsilon$ should be calculated in
the PG coordinate and one then has $\varepsilon=r_{\varepsilon}.$ If one has
$\varepsilon=\alpha m_{p}$, the atmosphere entropy around the horizon becomes%
\begin{equation}
S\sim\frac{A\kappa}{720\pi\alpha m_{p}}+s_{0}^{0}\ln\kappa m_{p}+%
%TCIMACRO{\dsum \limits_{l=1}^{\infty}}%
%BeginExpansion
{\displaystyle\sum\limits_{l=1}^{\infty}}
%EndExpansion
\frac{s_{l}}{\left(  m_{p}\kappa\right)  ^{l}}+\text{Finite terms as }%
m_{p}\kappa\rightarrow0,\label{eq:entropy-ff}%
\end{equation}
where we define $s_{l}=%
%TCIMACRO{\dsum \limits_{i=\max\left\{  1,l-1\right\}  }^{\infty}}%
%BeginExpansion
{\displaystyle\sum\limits_{i=\max\left\{  1,l-1\right\}  }^{\infty}}
%EndExpansion
\frac{s_{l+2i}^{i}}{\left(  2\pi\right)  ^{2i}\alpha^{l+2i}}$. Moreover, eqn.
$\left(  \ref{eq:entropy-ff}\right)  $ might suggest that effects of the MDR
on the atmosphere entropy is nonperturbative in this case. Alternatively,
inspired by the static scenario, one could choose $r_{\varepsilon}$ such that
the higher order terms in eqn. $\left(  \ref{eq:divergent-4D}\right)  $ have
the same order of divergence as $\frac{1}{\kappa r_{\varepsilon}}$. Here, we
could have
\begin{equation}
r_{\varepsilon}=\alpha m_{p}^{\frac{2}{3}}\kappa^{-\frac{1}{3},}%
\end{equation}
where $\alpha$ is some constant. In this case, the atmosphere entropy becomes%
\begin{equation}
S\sim\frac{\tilde{s}_{0}}{\alpha\left(  m_{p}\kappa\right)  ^{\frac{2}{3}}%
}+\frac{2s_{0}^{0}}{3}\ln m_{p}\kappa+\text{Finite terms as }m_{p}%
\kappa\rightarrow0,\label{eq:entropy-ff-4d}%
\end{equation}
where we define%
\begin{equation}
\tilde{s}_{0}=s_{1}^{0}\left[  1+%
%TCIMACRO{\dsum \limits_{i=1}^{\infty}}%
%BeginExpansion
{\displaystyle\sum\limits_{i=1}^{\infty}}
%EndExpansion
\frac{s_{3i+1}^{i}}{s_{1}^{0}}\left(  2\pi\alpha^{\frac{3}{2}}\right)
^{-2i}\right]  .\label{eq:s-tidal}%
\end{equation}
For a Schwarzschild black hole, the terms in the square bracket in eqn.
$\left(  \ref{eq:s-tidal}\right)  $ dose not depend on the black hole's
properties and the leading divergent term in eqn. $\left(
\ref{eq:entropy-ff-4d}\right)  $ scales with $A\kappa^{\frac{4}{3}}%
m_{p}^{-\frac{2}{3}}$. In \cite{EBWM-Jacobson:2007jx}, the authors calculated
the black hole horizon entanglement entropy for a massless scalar field with
the MDR imposed in a free-fall frame. With the sub- or super-luminal
dispersion with index $n$, they found that the entanglement entropy scales as
$A\kappa^{1+\frac{1}{n}}m_{p}^{-1+\frac{1}{n}}$.

Following the argument proposed in \cite{EBWM-AmelinoCamelia:2004xx}, the
authors in \cite{EBWM-AmelinoCamelia:2005ik} obtained modified relations
between the mass of a Schwarzschild black hole and its temperature and
entropy. The argument connecting a MDR and some modifications of the entropy
of black holes is formulated in a scheme of analysis first introduced by
Bekenstein\cite{EBWM-Bekenstein:1973ur}. In fact, for the MDR in eqn. $\left(
\ref{eq:MDR}\right)  $, the modified temperature of the black hole was given
by%
\begin{equation}
T_{eff}=\frac{1}{4\pi}F\left(  \frac{m_{p}^{2}}{2M}\right)  ,
\end{equation}
where $M$ is the mass of the black hole. The first law of black hole
thermodynamics $dS_{B}=\frac{dM}{T}$ and eqn. $\left(  \ref{eq:F(p)}\right)  $
lead to the modified entropy of the black hole%
\begin{equation}
S_{B}=\frac{A}{4m_{p}^{2}}+2\pi C_{1}\ln\kappa m_{p}+\text{Finite terms as
}m_{p}\kappa\rightarrow0, \label{eq:entropy-BH}%
\end{equation}
where $A=16\pi M^{2}$ and $\kappa=\frac{1}{4M}.$ It is interesting to note
that the modifications of the entropy of black holes in eqn. $\left(
\ref{eq:entropy-BH}\right)  $ are finite as $m_{p}\kappa\rightarrow0$ except
the logarithmic term. If one wants the same story for the atmosphere entropy
obtained in eqn. $\left(  \ref{eq:divergent-4D}\right)  $, one could have
$r_{\varepsilon}=\alpha\kappa^{\delta-1}m_{p}^{\delta}$ for some constant
$\alpha$ where $0<\delta\leq\frac{2}{3}$ in the static scenario and
$0<\delta\leq\frac{1}{2}$ in the free-fall scenario. Hence, the atmosphere
entropy becomes%
\begin{equation}
S\sim\frac{s_{1}^{0}}{\alpha\left(  \kappa m_{p}\right)  ^{\delta}}+s_{0}%
^{0}\delta\ln\kappa m_{p}+\text{Finite terms as }m_{p}\kappa\rightarrow0,
\end{equation}
where the leading divergent term scales with $A\kappa^{2-\delta}m_{p}%
^{-\delta}$ for a Schwarzschild black hole. The coefficient of the logarithmic
term in eqn. $\left(  \ref{eq:entropy-BH}\right)  $ depends on $C_{1}$, a
coefficient of the MDR. However, we show that the coefficient of the
subleading logarithmic term in the atmosphere entropy is irrelevant to
coefficients of the MDR. It only depends on the position of the wall,
$r_{\varepsilon}$ and the properties of the black hole.

For a 2D black hole with the Schwarzschild-like coordinate%
\begin{equation}
ds^{2}=f\left(  r\right)  dt^{2}-\frac{dr^{2}}{f\left(  r\right)  },
\end{equation}
the atmosphere entropy of a massless scalar can also be presented in the form
of a Laurent series with respect to $r_{\varepsilon}$%
\begin{equation}
S\sim s_{0}^{0}\ln\kappa r_{\varepsilon}+%
%TCIMACRO{\dsum \limits_{i=1}^{\infty}}%
%BeginExpansion
{\displaystyle\sum\limits_{i=1}^{\infty}}
%EndExpansion
\frac{T_{0}^{2i}}{m_{p}^{2i}}\left(
%TCIMACRO{\dsum \limits_{j=1}^{\delta_{i}}}%
%BeginExpansion
{\displaystyle\sum\limits_{j=1}^{\delta_{i}}}
%EndExpansion
s_{j}^{i}\left(  \kappa r_{\varepsilon}\right)  ^{-j}+s_{0}^{i}\ln\kappa
r_{\varepsilon}\right)  , \label{eq:entropy-2d-ff-r}%
\end{equation}
where $\delta_{i}=2i$ in the static scenario and $\delta_{i}=3i$ in the
free-fall scenario. From eqns. $\left(  \ref{eq:entropy-2d-ff}\right)  $ and
$\left(  \ref{eq:entropy-sf-4d}\right)  $, one has $s_{0}^{0}=-\frac{1}{12}$.
In the static scenario, if we assume that the proper length $\varepsilon$ is
measured along a static time slice and $\varepsilon=\alpha m_{p}$, the
atmosphere entropy of a massless scalar becomes
\begin{equation}
S\sim-\frac{1}{6}\ln\kappa m_{p}+\text{Finite terms as }\kappa m_{p}%
\rightarrow0,
\end{equation}
where the same leading logarithmic term was also obtained in
\cite{EBWM-Mann:1990fk}\ for the scenario without the MDR. In the free-fall
scenario, if the proper length $\varepsilon$ is assumed to be measured on a
time slice orthogonal to the free-fall world lines and we let $\varepsilon
=\alpha m_{p}$, the atmosphere entropy of a massless scalar becomes%
\begin{equation}
S\sim-\frac{1}{12}\ln\kappa m_{p}+\frac{1}{4\pi^{2}}\sum_{n=1}^{\infty}%
\frac{s_{n}}{\left(  \kappa m_{p}\right)  ^{n}}+\text{Finite terms as }\kappa
m_{p}\rightarrow0,
\end{equation}
where $s_{n}$ can be determined by eqn. $\left(  \ref{eq:entropy-2d-ff}%
\right)  $. If we want that the modifications of the entropy of black holes in
eqn. $\left(  \ref{eq:entropy-2d-ff-r}\right)  $ are finite, we could have
$r_{\varepsilon}=\alpha\kappa^{\delta-1}m_{p}^{\delta}$ for some constant
$\alpha$ where $0<\delta\leq\frac{2}{3}$ for some constant $\alpha$. The
entropy then becomes%
\begin{equation}
S\sim-\frac{\delta}{12}\ln\kappa m_{p}+\text{Finite terms as }\kappa
m_{p}\rightarrow0.
\end{equation}

\section{Black Hole Evaporation}

\label{Sec:BHE}

In the section, we discuss the MDR effects on the evaporations of a
Schwarzschild black hole to $\mathcal{O}\left(  \frac{T_{0}^{2}}{m_{p}^{2}%
}\right)  $ in the free-fall scenario. For simplicity, we assume that the
emitted particles are massless. In \cite{BHE-Page:1976df}, Page counted the
number of modes per frequency interval with periodic boundary conditions in a
large container around the black hole and divided it by the time it takes a
particle to cross the container. He then related the expected number emitted
per mode $n_{\omega,l}$ to the average emission rate per frequency interval
$\frac{dn_{\omega,l}}{dt}$ by%
\begin{equation}
\frac{dn_{\omega,l}}{dt}=n_{\omega,l}\frac{d\omega}{2\pi\hbar},\label{eq:DnDt}%
\end{equation}
for each mode and frequency interval $\left(  \omega,\omega+d\omega\right)  $.
Following the same argument, we find that in the MDR case
\begin{equation}
\frac{dn_{\omega,l}}{dt}=n_{\omega,l}\frac{\partial\omega}{\partial p_{r}%
}\frac{dp_{r}}{2\pi\hbar}=n_{\omega,l}\frac{d\omega}{2\pi\hbar}%
,\label{eq:DnDt-MDR}%
\end{equation}
where $\frac{\partial\omega}{\partial p_{r}}$ is the radial velocity of the
particle and the number of modes between the wavevector interval $\left(
p_{r},p_{r}+dp_{r}\right)  $ is $\frac{dp_{r}}{2\pi\hbar}$. Since each
particle carries off the energy $\omega$, the total luminosity is obtained
from $\frac{dn_{\omega,l}}{dt}$ by multiplying by the energy $\omega$ and
summing up over all energy $\omega$ and $l$,%
\begin{equation}
L=%
%TCIMACRO{\dsum \limits_{l=0}}%
%BeginExpansion
{\displaystyle\sum\limits_{l=0}}
%EndExpansion
\left(  2l+1\right)  \int\omega n_{\omega,l}\frac{d\omega}{2\pi\hbar}.
\end{equation}
However, some of the radiation emitted by the horizon might not be able to
reach the asymptotic region. Before the radiation reaches the distant
observer, they must pass the curved spacetime around the black hole horizon,
which plays the role of a potential barrier. This effect on $L$ can be
described by a greybody factor from the scattering coefficients of the black
hole. Actually, the greybody factor is given by $\left\vert T_{l}\left(
\omega\right)  \right\vert ^{2}$, where $T_{l}\left(  \omega\right)  $
represents the transmission coefficient of the black hole barrier which in
general depends on the energy $\omega$ and angular momentum $l$ of the
particle. Taking the greybody factor into account, we find for the total
luminosity%
\begin{equation}
L=%
%TCIMACRO{\dsum \limits_{l=0}}%
%BeginExpansion
{\displaystyle\sum\limits_{l=0}}
%EndExpansion
\left(  2l+1\right)  \int\left\vert T_{l}\left(  \omega\right)  \right\vert
^{2}\omega n_{\omega,l}\frac{d\omega}{2\pi\hbar}\text{\thinspace
}.\label{eq:Luminosity}%
\end{equation}
The relevant radiation usually have the energy of order $\hbar M^{-1}$ for a
black hole with the mass $M,$ one hence needs to use the wave equations given
in the appendix of \cite{SF} to compute $\left\vert T_{l}\left(
\omega\right)  \right\vert ^{2}$ accurately. However, solving the wave
equations for $\left\vert T_{l}\left(  \omega\right)  \right\vert ^{2}$ could
be very complicated. On the other hand, we can use the geometric optics
approximation to estimate $\left\vert T_{l}\left(  \omega\right)  \right\vert
^{2}$. In the geometric optics approximation, we assume $\omega\gg M$ and high
energy waves will be absorbed unless they are aimed away from the black hole.
Hence $\left\vert T_{l}\left(  \omega\right)  \right\vert ^{2}=1$ for all the
classically allowed energy $\omega$ and the angular momentum $l$, while
$\left\vert T_{i}\left(  E\right)  \right\vert ^{2}=0$ otherwise. For the
usual dispersion relation, the well-known Stefan's law for black holes is
obtained in this approximation.

To find the classically allowed angular momentum $l$ with fixed value of
energy $\omega$, we consider eqn. $\left(  \ref{eq:Deformed-HJ}\right)  $ for
a massless particle in the Schwarzschild black hole with the mass $M$. Solving
eqn. $\left(  \ref{eq:Deformed-HJ}\right)  $ for $\lambda$ to $\mathcal{O}%
\left(  m_{p}^{-2}\right)  $ gives
\begin{equation}
\frac{\lambda}{r^{2}}=\left(  \omega-vp_{r}\right)  ^{2}-\frac{2C_{1}}%
{m_{p}^{2}}\left(  \omega-vp_{r}\right)  ^{4}-p_{r}^{2},
\label{eq:Deformed HJ-4D}%
\end{equation}
where we have $\lambda=\left(  l+\frac{1}{2}\right)  ^{2}\hbar^{2}$, $v\left(
r\right)  =-\sqrt{\frac{2M}{r}}$ and $C_{1}$ is given in eqn. $\left(
\ref{eq:MDR}\right)  $. In the geometric optics approximation, $p_{r}$ is
always a real number. In the non-dipsersive case $\left(  m_{p}\rightarrow
\infty\right)  $, the maximum of the RHS of eqn. $\left(
\ref{eq:Deformed HJ-4D}\right)  $ is $\frac{\omega^{2}}{1-\frac{2M}{r}}$.
Thus, one has
\begin{equation}
\lambda\leq\frac{r^{2}\omega^{2}}{1-\frac{2M}{r}},
\end{equation}
where the RHS has a minimum at $r_{\min}=3M$, which is $27M^{2}\omega^{2}$. If
the particles overcome the angular momentum barrier and get absorbed by the
black hole, one must have $\lambda\leq27M^{2}\omega^{2}$. In the geometric
optics approximation, the Schwarzschild black hole is just like a black sphere
of radius $R=3^{3/2}M$\cite{Wald:1984rg}. When the second term in the RHS of
eqn. $\left(  \ref{eq:Deformed HJ-4D}\right)  $ is included, the maximum of
the RHS is shifted to
\begin{equation}
\Omega\equiv\frac{\omega^{2}}{1-\frac{2M}{r}}-\frac{2C_{1}\omega^{2}}%
{m_{p}^{2}}\frac{\omega^{2}}{\left(  1-\frac{2M}{r}\right)  ^{4}}.
\end{equation}
The minimum of $r^{2}\Omega$ is then shifted to%
\begin{equation}
\lambda_{\max}\equiv27M^{2}\omega^{2}\left(  1-\frac{54C_{2}\omega^{2}}%
{m_{p}^{2}}\right)  .
\end{equation}
Therefore, the particle must have $\lambda\leq\lambda_{\max}$ to get absorbed
by the black hole. The eqn. $\left(  \ref{eq:Luminosity}\right)  $ then
becomes%
\begin{equation}
L=\int\frac{\omega d\omega}{2\pi\hbar^{3}}\int_{0}^{\lambda_{\max}}n\left(
\frac{\omega\left(  1+\Delta\right)  }{T_{0}}\right)  d\lambda,
\label{eq:L-Delta}%
\end{equation}
where eqn. $\left(  \ref{eq:Number}\right)  $ for $n_{\omega,l}$ is used. For
a massless particle in the Schwarzschild black hole, eqn. $\left(
\ref{eq:Delta-ff}\right)  $ gives%
\begin{equation}
\Delta=-\frac{C_{1}}{2m_{p}^{2}}\left(  20\omega^{2}-\frac{3\lambda}{4M^{2}%
}\right)  . \label{eq:delta-ff-Schwarzchild}%
\end{equation}
Expanding $n\left(  \frac{\omega\left(  1+\Delta\right)  }{T_{0}}\right)  $ to
$\mathcal{O}\left(  m_{p}^{-2}\right)  $ and then integrating eqn. $\left(
\ref{eq:L-Delta}\right)  $, we find for the emission of $n_{s}$ species of
massless scalars and $n_{f}$ species of massless spin-$1/2$ fermions that the
total luminosity is
\begin{equation}
L=\frac{9m_{p}^{2}}{40960M^{2}}\left[  \left(  n_{s}+\frac{7}{4}n_{f}\right)
-C_{1}\left(  0.73n_{s}+1.41n_{f}\right)  \frac{m_{p}^{2}}{M^{2}}%
+\mathcal{O}\left(  \frac{m_{p}^{3}}{M^{3}}\right)  \right]  . \label{eq:L-ff}%
\end{equation}

To make a comparison with the static scenario, the results in \cite{SF} are
given below for the form of MDR $\left(  \ref{eq:MDR}\right)  .$ In the static
scenario, the correction to the Hawking temperature is
\begin{equation}
\Delta=-\frac{C_{1}}{2m_{p}^{2}}\left(  4\omega^{2}+\frac{\lambda}{4M^{2}%
}\right)  ,
\end{equation}
the maximum of the angular momentum is%
\begin{equation}
\lambda_{\max}=27M^{2}\omega^{2}\left(  1-\frac{6C_{1}\omega^{2}}{m_{p}^{2}%
}\right)  ,
\end{equation}
and the total luminosity is%
\begin{equation}
L=\frac{9m_{p}^{2}}{40960\pi M^{2}}\left[  \left(  n_{s}+\frac{7}{4}%
n_{f}\right)  +C_{1}\left(  0.48n_{s}+0.92n_{f}\right)  \frac{m_{p}^{2}}%
{M^{2}}+\mathcal{O}\left(  \frac{m_{p}^{3}}{M^{3}}\right)  \right]
.\label{eq:L-sf}%
\end{equation}
Note that the sign in front of $C_{1}$ in eqn. $\left(  \ref{eq:L-ff}\right)
$ is different from that in eqn. $\left(  \ref{eq:L-sf}\right)  .$ For the
sub-luminal dispersion relation with $C_{1}<0$, it means that the total
luminosity increases due to the MDR effects in the free-fall scenario while it
decreases in the static scenario. In the geometric optics approximation, the
black hole can be described as a black sphere for absorbing particles. The
total luminosity are determined by the radius of the black sphere $R$ and the
temperature of the black hole $T$. Note that one has $R=\sqrt{\frac
{\lambda_{\max}}{\omega^{2}}}$ and $T_{eff}\approx T_{0}\left(  1-\Delta
\right)  $. In the static scenarios, the MDR effects increase the radius of
the black sphere while they decrease the temperature of the black hole. The
competition between the increased radius and the decreased temperature
determines whether the luminosity would increase or decrease. It appears from
eqn.$\left(  \ref{eq:L-sf}\right)  $ that the effects of decreased temperature
win the competition. In the free-fall scenario, the MDR effects also increase
the radius of the black sphere. Due to the minus sign in front of
$\frac{3\lambda}{4M^{2}}$ in eqn. $\left(  \ref{eq:delta-ff-Schwarzchild}%
\right)  $, the temperature of the black hole increases for $\lambda>\frac
{80}{3}M^{2}\omega^{2}$ and decreases for $\lambda<\frac{80}{3}M^{2}\omega
^{2}$, but more slowly than in the static scenario. As a result, the effects
of increased radius win the competition and hence the luminosity increases.
The opposite story happens to the super-luminal case with $C_{1}>0$.

\section{Conclusion}

\label{Sec:Con}

In this paper, we used the Hamilton-Jacobi method to calculate tunneling rates
of radiations across the horizon and the effective Hawking temperatures in the
dispersive models with the free-fall preferred frame.\ After the near horizon
spectrum of radiations was obtained, the thermal entropy of radiations near
the horizon and luminosity of the black hole were computed. Our main results are:

\begin{itemize}
\item In section \ref{Sec:DHJM}, we first derived the deformed Hamilton-Jacobi
equations in the dispersive models with the free-fall preferred frame. The
deformed Hamilton-Jacobi equations were then solved for $\partial_{r}I$ and
the imaginary part of $I$ was obtained. The corrections to the Hawking
temperature were calculated for massive and charged particles to
$\mathcal{O}\left(  m_{p}^{-2}\right)  $ and neutral and massless particles
with $\lambda=0$ to all orders, respectively. It was found that corrections
were suppressed by $m_{p}$.

\item In section \ref{Sec:EBWM}, we used the brick wall model to compute the
thermal entropy of a massless scalar field near the horizon of a 4D
spherically symmetric black hole and a 2D one. For a 4D black hole, the
entropy near the horizon have been calculated to $\mathcal{O}\left(
m_{p}^{-2}\right)  $ and could be written in the form of eqn. $\left(
\ref{eq:divergent-4D}\right)  $. The entropy became divergent as the wall
approached the horizon. Various choices of the proper distance between the
wall and the horizon and the corresponding entropies have been discussed. For
a 2D black hole, entropies in the static and free-fall scenarios have been
calculated to all orders. The leading divergent terms were logarithmic.
Nevertheless, their coefficients depended on choices of the proper distance
between the wall and the horizon.

\item In section \ref{Sec:BHE}, we calculated luminosities of a Schwarzschild
black hole with the mass $M\gg m_{p}$. We used the geometric optics
approximation to estimate the effects of scattering off the background.
Comparison between the static scenario and the free-fall one has been given there.
\end{itemize}

\begin{acknowledgments}
We are grateful to Houwen Wu and Zheng Sun for useful discussions. This work
is supported in part by NSFC (Grant No. 11005016, 11175039 and 11375121) and
the Fundamental Research Funds for the Central Universities.
\end{acknowledgments}

\end{document}